\documentclass[twoside,twocolumn,9pt]{article}
\usepackage{extsizes}
\usepackage[super,sort&compress,comma]{natbib} 
\usepackage[version=3]{mhchem}
\usepackage[left=1.5cm, right=1.5cm, top=1.785cm, bottom=2.0cm]{geometry}
\usepackage{balance}
\usepackage{mathptmx}
\usepackage{sectsty}
\usepackage{graphicx} 
\usepackage{lastpage}
\usepackage{booktabs} 
\usepackage{multirow} 
\usepackage{hyperref}
\usepackage{orcidlink}
\usepackage{tabularx}
\usepackage{array}
\usepackage[format=plain,justification=justified,singlelinecheck=false,font={stretch=1.125,small,sf},labelfont=bf,labelsep=space]{caption}
\usepackage{float}
\usepackage{fancyhdr}
\usepackage{fnpos}
\usepackage[english]{babel}
\addto{\captionsenglish}{%
  
}
\usepackage{array}
\usepackage{droidsans}
\usepackage{charter}
\usepackage[T1]{fontenc}
\usepackage[usenames,dvipsnames]{xcolor}
\usepackage{setspace}
\usepackage[compact]{titlesec}

\usepackage{epstopdf}
\usepackage{comment}

\definecolor{cream}{RGB}{222,217,201}

\begin{document}

\pagestyle{fancy}
\thispagestyle{plain}
\fancypagestyle{plain}{
\renewcommand{\headrulewidth}{0pt}
}

\makeFNbottom
\makeatletter
\renewcommand\LARGE{\@setfontsize\LARGE{15pt}{17}}
\renewcommand\Large{\@setfontsize\Large{12pt}{14}}
\renewcommand\large{\@setfontsize\large{10pt}{12}}
\renewcommand\footnotesize{\@setfontsize\footnotesize{7pt}{10}}
\makeatother

\renewcommand{\thefootnote}{\fnsymbol{footnote}}
\renewcommand\footnoterule{\vspace*{1pt}%
\color{cream}\hrule width 3.5in height 0.4pt \color{black}\vspace*{5pt}} 
\setcounter{secnumdepth}{5}

\makeatletter 
\renewcommand\@biblabel[1]{#1}            
\renewcommand\@makefntext[1]%
{\noindent\makebox[0pt][r]{\@thefnmark\,}#1}
\makeatother 
\renewcommand{\figurename}{\small{Fig.}~}
\sectionfont{\sffamily\Large}
\subsectionfont{\normalsize}
\subsubsectionfont{\bf}
\setstretch{1.125} 
\setlength{\skip\footins}{0.8cm}
\setlength{\footnotesep}{0.25cm}
\setlength{\jot}{10pt}
\titlespacing*{\section}{0pt}{4pt}{4pt}
\titlespacing*{\subsection}{0pt}{15pt}{1pt}

\fancyfoot{}
\fancyfoot[LO,RE]{\vspace{-7.1pt}\includegraphics[height=9pt]{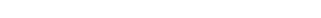}}
\fancyfoot[CO]{\vspace{-7.1pt}\hspace{13.2cm}\includegraphics{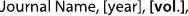}}
\fancyfoot[CE]{\vspace{-7.2pt}\hspace{-14.2cm}\includegraphics{head_foot/RF}}
\fancyfoot[RO]{\footnotesize{\sffamily{1--\pageref{LastPage} ~\textbar  \hspace{2pt}\thepage}}}
\fancyfoot[LE]{\footnotesize{\sffamily{\thepage~\textbar\hspace{3.45cm} 1--\pageref{LastPage}}}}
\fancyhead{}
\renewcommand{\headrulewidth}{0pt} 
\renewcommand{\footrulewidth}{0pt}
\setlength{\arrayrulewidth}{1pt}
\setlength{\columnsep}{6.5mm}
\setlength\bibsep{1pt}

\makeatletter 
\newlength{\figrulesep} 
\setlength{\figrulesep}{0.5\textfloatsep} 

\newcommand{\topfigrule}{\vspace*{-1pt}%
\noindent{\color{cream}\rule[-\figrulesep]{\columnwidth}{1.5pt}} }

\newcommand{\botfigrule}{\vspace*{-2pt}%
\noindent{\color{cream}\rule[\figrulesep]{\columnwidth}{1.5pt}} }

\newcommand{\dblfigrule}{\vspace*{-1pt}%
\noindent{\color{cream}\rule[-\figrulesep]{\textwidth}{1.5pt}} }

\makeatother

\twocolumn[
  \begin{@twocolumnfalse}
{\includegraphics[height=30pt]{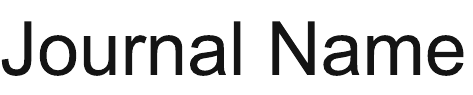}\hfill\raisebox{0pt}[0pt][0pt]{\includegraphics[height=55pt]{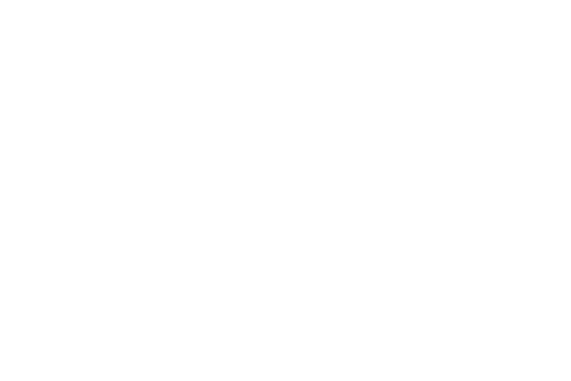}}\\[1ex]
\includegraphics[width=18.5cm]{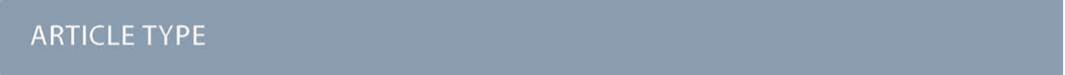}}\par
\vspace{1em}
\sffamily
\begin{tabular}{m{4.5cm} p{13.5cm} }

\includegraphics{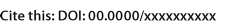} & \noindent\LARGE{\textbf{Linker Functionalization and pH Tuning Enhance Solar-Driven Catalytic CO$_2$ Reduction in MOF-5$^\dag$}} \\
\vspace{0.3cm} & \vspace{0.3cm} \\

 & \noindent\large{Julia Santana-Andreo\orcidlink{0000-0003-1640-327X},$^{\ast}$\textit{$^{a,b}$} Joshua Edzards\orcidlink{0009-0002-3347-2576},\textit{$^{a,b}$} Surender Kumar\orcidlink{0009-0000-3072-5633},$^{\ast}$\textit{$^{a}$} and Caterina Cocchi\orcidlink{0000-0002-9243-9461}$^{\ast}$\textit{$^{a,b,c}$}} \\

\includegraphics{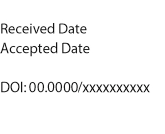} & \noindent\normalsize{The wide band gap of metal-organic framework (MOF) 5 constrains its use in photocatalytic carbon dioxide (CO$_2$) reduction despite its high porosity and favorable mass-transport properties. Adopting a state-of-the-art first-principles approach, we systematically investigate the effects of volumetric strain, metal-node substitution, linker functionalization, and pH control as knobs to improve the CO$_2$ photocatalytic ability of MOF-5. Strain and metal-node substitution negligibly affect the gap, whereas linker functionalization narrows it into the visible range via in-gap states while preserving reduction-side alignment at pH~=~0. The resulting reduction energetics are strongly sensitive to both linker functionalization and pH. Halogenated and hydroxylated frameworks provide access primarily to HCOOH, CO, and HCHO under alkaline conditions, principally with the Mg and Zn nodes, whereas COOH functionalization offers the broadest thermodynamic accessibility across the full CO$_2$ reduction sequence. NH$_2$ retains thermodynamic feasibility for all target reduction pathways but with larger overpotentials, while NO$_2$ generally yields unfavorable reduction energetics. Crucially, within the COOH series, the choice of the metal node tunes the fundamental gap by over 1~eV with only minor changes in the reduction overpotentials, placing Sr- and Ba-based architectures as the most favorable ones for broad thermodynamic product accessibility with visible-light excitation.  Linker functionalization substantially reduces the spatial overlap of the frontier states, which is expected to favor photoinduced charge separation. Taken together, these results establish linker functionalization as a design lever that, combined with the well-known pH dependence of proton-coupled CO$_2$ reduction, enables the independent tuning of light absorption and CO$_2$-reduction energetics in MOF-5, establishing a rational and viable route for designing efficient MOF-based photocatalysts. } \\

\end{tabular}

 \end{@twocolumnfalse} \vspace{0.6cm}

  ]

\renewcommand*\rmdefault{bch}\normalfont\upshape
\rmfamily
\section*{}
\vspace{-1cm}


\footnotetext{\textit{$^{a}$~Institut für Festk\"orpertheorie und -Optik, Friedrich-Schiller-Universit\"at Jena, 07743 Jena, Germany}}
\footnotetext{\textit{$^{b}$~Carl von Ossietzky Universit\"at Oldenburg, Institute of Physics, 26129 Oldenburg, Germany}}
\footnotetext{\textit{$^{c}$~Abbe Center of Photonics, Friedrich-Schiller-Universit\"at Jena, 07745, Jena, Germany}}
\footnotetext{$^*$Email: julia.santana.andreo@uni-jena.de, surendermohinder@gmail.com, caterina.cocchi@uni-jena.de}

\footnotetext{\dag~Supplementary Information available: See DOI: 00.0000/00000000.}



\section{INTRODUCTION}
The continuing rise in atmospheric carbon dioxide (CO$_2$) concentrations is a major driver of global climate change and associated environmental impacts~\cite{Lacis2010}. This challenge motivates the development of renewable, low-cost, and sustainable strategies for CO$_2$ capture, separation, and utilization, including chemical and physical absorption, membrane separation, and adsorption-based processes~\cite{Aaron2005,CLAUSSE2011,LEUNG2014,Zhang2020,SANNI2022}. 
In particular, photocatalytic CO$_2$ reduction is an attractive route for transforming CO$_2$ into fuels and value-added chemicals such as methane and methanol~\cite{Fang2023,Xie2026,Liu2026,Jung2026}. 
However, achieving high conversion yields depends heavily on the intrinsic photophysical properties of the catalyst \cite{Fang2023}.
Efficient photocatalysis requires photogenerated electron--hole pairs to separate and migrate to reactive sites before recombination \cite{torkashvand2026}, while preserving the thermodynamic viability of relevant redox reactions~\cite{Bertuletti2026,Zuo2026}. Effective light harvesting and charge separation must therefore be combined with chemically active and accessible sites that promote reactant adsorption and activation~\cite{Zuo2026}.

Metal--organic frameworks (MOFs) are well-established platforms for photocatalysis thanks to their high surface areas, permanent porosity, and chemically tunable composition~\cite{wu2018,wu2019,Liu2026}. Their ordered porous structures facilitate adsorption, diffusion, and activation of gas molecules and have motivated applications in gas storage, sensing, and catalysis~\cite{Kaye2007,LI2009,Zhao2013,YANG2012,wu2018}. Among experimentally accessible frameworks, MOF-5~\cite{Li1999} represents one of the most extensively investigated  systems~\cite{LI2009,Eddaoudi2002,fabrizio2022,Furukawa2013,Hafizovic2007}.
However, despite structural robustness, its applicability as a solar photocatalyst is limited by its wide band gap~\cite{Alvaro2007, fabrizio2022} compatible with absorption of ultraviolet radiation~\cite{fabrizio2022,Edzards2024} and its reduced charge-carrier separation efficiency~\cite{Alvaro2007,Nasalevich2014}.

\begin{figure*}[!ht]
\centering
\includegraphics[width=0.9\textwidth]{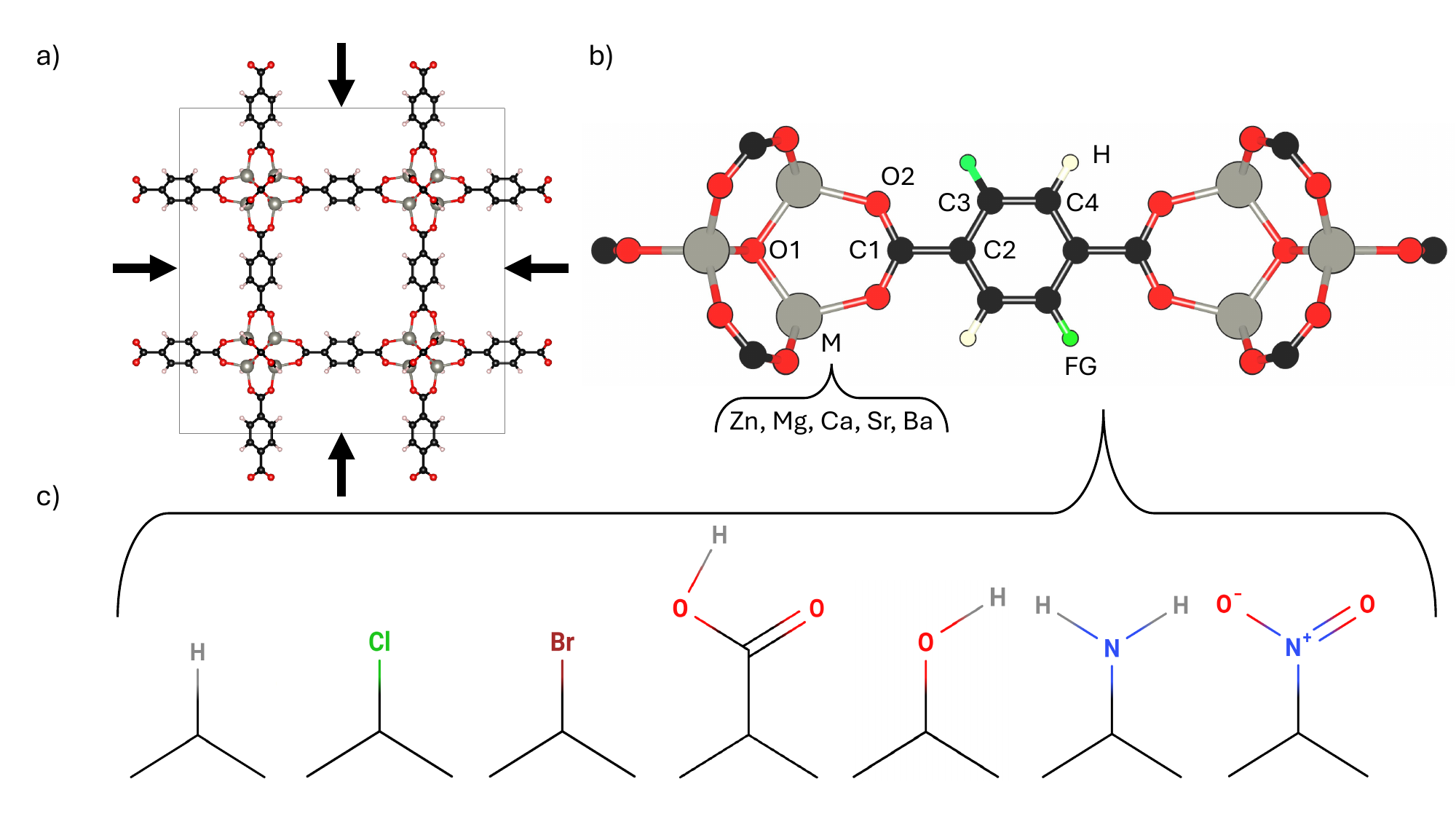}
\caption{a) MOF-5 in its cubic conventional cell under isotropic strain, illustrated by black arrows in the case of compression; b) local structural motif including the metal node M, depicted in gray, and the BDC linker, with carbon atoms in black, hydrogen in white, oxygen in red, and functional groups (FG) in green; c) schematic visualization of the considered FGs, including H, Cl, Br,  COOH, OH, NH$_2$, and NO$_2$.}
\label{fig:mof5-scheme}
\end{figure*}

To overcome these issues, exploiting the modular structure of MOF-5 via metal-ion substitution and linker functionalization represents a viable and efficient route~\cite{butler2014electronic} to tune its electronic structure~\cite{YANG2012,Hausdorf2010,Botas2010,YANG2013,Sugamata2021,syzgantseva2019}, improve its light harvesting ability~\cite{Hendon2013}, and enhance charge-carrier mobilities therein~\cite{calvo2020,zojer2021,Winkler2020}. In particular, linker functionalization can induce substantial structural rearrangements and introduce electronic states within the band gap~\cite{Edzards2024}, providing a direct strategy for tuning both light absorption and the energies and spatial character of the frontier energy states.

In this work, we use density-functional theory (DFT) to evaluate how structural and chemical modifications can tailor MOF-5 for solar-driven CO$_2$ photocatalysis. Specifically, we compare the effects of lattice strain, metal-node substitution, and linker functionalization on electronic gaps, band alignments, and frontier orbital spatial separation, identifying the latter as the primary driver for performance optimization. The chemical modification of the ligands via molecular substituents introduce in-gap states that markedly reduce the electronic gap and the spatial overlap of frontier states while preserving favorable band alignment. By integrating these electronic structure insights with thermodynamic modeling, we map how the Nernstian pH dependence of the proton-coupled reduction couples translates into framework-specific pH windows, showing that the thermodynamic accessibility of the different reduction products is governed jointly by the framework design and operating conditions.

\section{COMPUTATIONAL FRAMEWORK}\label{sec:method}

The DFT~\cite{hohn+64pr,kohn+65pr} calculations presented in this work were performed using \texttt{CP2K}~\cite{kuhn+20jcp} (version 2024.1), employing \texttt{TZV2P-MOLOPT-MGGA-GTH} Gaussian basis sets~\cite{vand-hutt07jcp} of triple-$\zeta$ quality with two polarization functions. Calculations employed the Gaussian and plane-wave formalism~\cite{lipp+97mp} implemented in \texttt{QUICKSTEP}\cite{vand+05cpc} in which the electronic states are represented by localized Gaussian basis functions while the electronic density is mapped onto an auxiliary plane-wave grid for the solution of the Poisson equation.  Core electrons were described by Goedecker--Teter--Hutter pseudopotentials~\cite{goed+96prb}, while exchange--correlation effects were accounted for by adopting the r$^2$SCAN functional~\cite{furn+20jpcl}, including dispersion interactions through the D3 correction~\cite{grim+10jcp}. This choice results from systematic benchmarks for MOF-5 and its derivatives~\cite{edzards2025,SantanaAndreo2026}. For all calculations, a real-space integration grid defined by plane-wave cutoff 20.4~keV and Gaussian-type orbitals cutoff 2.7~keV was employed. The self-consistent field convergence threshold was set to $5\times10^{-7}$, and geometries were optimized until the maximum interatomic force fell below $2.6\times10^{-2}$~eV/\AA{}.

To efficiently model structurally and chemically modified MOF-5 while preserving its symmetry, two unit cell representations were employed~\cite{SantanaAndreo2026}. The strain and metal-substitution series were treated in the conventional cubic cell containing 424 atoms. This setting keeps the cubic axes explicit, allowing an unambiguous application of strain along all three crystallographic directions while preserving the symmetry equivalence of the eight M$_4$O clusters (M = Mg, Ca, Sr, Ba). Owing to the large lattice parameter ($a_{\mathrm{cub}} = 25.85$~\AA{} in good agreement with the experimental range 25.83--25.89~\AA{}~\cite{Eddaoudi2002,Hafizovic2007}) and correspondingly small Brillouin zone, these calculations were sampled at $\Gamma$ only. On the other hand, linker-functionalized frameworks were modelled in the primitive rhombohedral cell (106 atoms, see Table~S1), since this chemical modification inherently lowers the cubic symmetry. The corresponding Brillouin zone was sampled with a Monkhorst--Pack $4\times4\times4$ $k$-point mesh~\cite{monk+76prb}, yielding a denser sampling at one quarter of the computational cost compared to the conventional cell.  Benchmark calculations on pristine MOF-5 confirmed excellent agreement between the two cell representations, with electronic gaps matching within 0.08~eV (3.83~eV \textit{vs.} 3.91~eV) and pore reference potentials within 0.04~eV. 

Absolute band-edge energies relative to the vacuum level were determined following the approach proposed by Butler et al.~\cite{butler2014}, using the electrostatic potential within the pore cavity of the framework as an internal reference~\cite{harnett2021}. Specifically, the potential was spherically averaged within a 2~\AA{} radius sphere centered at the pore midpoint, as implemented in the \texttt{MacroDensity} package~\cite{macrodensity}. The converged potential value was then used to align the energy levels to the vacuum scale.

All gaps reported in the following are Kohn--Sham gaps, defined as the
absolute difference between the energies of the valence-band maximum (VBM) and
the conduction-band minimum (CBM), and are referred to as fundamental gaps
throughout. For photocatalysis, the relevant quantity is instead the optical gap, which is lowered with respect to the fundamental gap by the exciton binding energy. In porous materials like MOFs, localized frontier states and weak dielectric screening yield tightly bound electron--hole pairs alongside substantial quasiparticle corrections~\cite{Kshirsagar2021}. Remarkably, these two effects partially cancel out when adopting advanced functionals like r$^2$SCAN~\cite{edzards2025}, so that Kohn--Sham gaps provide reasonable estimates of the absorption onset as shown for pristine MOF-5, the r$^2$SCAN gap lying close to the lower bound of the experimental range of 3.87--4.71~eV.

\section{RESULTS}
To establish baseline properties for structural and chemical tuning, pristine MOF-5 is evaluated as our reference framework. We systematically explore three complementary modifications: (i) isotropic volumetric strain ($-5.0$ to $+5.0$\%, Fig.~\ref{fig:mof5-scheme}a), (ii) metal-node substitution (Zn $\to$ Mg, Ca, Sr, Ba, Fig.~\ref{fig:mof5-scheme}b), and (iii) symmetric functionalization of the benzene-1,4-dicarboxylate (BDC) linker at the 2- and 5-positions with electron-donating (OH, NH$_2$) and electron-withdrawing (Br, Cl, COOH, NO$_2$) groups (Fig.~\ref{fig:mof5-scheme}c). 

The dynamical stability of pristine MOF-5  at the r$^2$SCAN+D3 level of theory was established in previous work~\cite{SantanaAndreo2026}. There, a shallow  $A_{2\mathrm{g}}$ soft mode in the high-symmetry $Fm\bar{3}m$ phase prompted the lower-symmetry $Fm\bar{3}$ configuration as our reference framework. Assessing the dynamical stability of metal-exchanged and particularly functionalized derivatives is extremely challenging using conventional perturbative approaches, due to the shallowness of the corresponding potential energy surface. However, since the underlying distortion is a collective node-linker deformation on a sub-meV energy scale, its impact on the frontier-level energies discussed below is expected to be negligible.
\begin{figure}
    \centering
    \includegraphics[width=\linewidth]{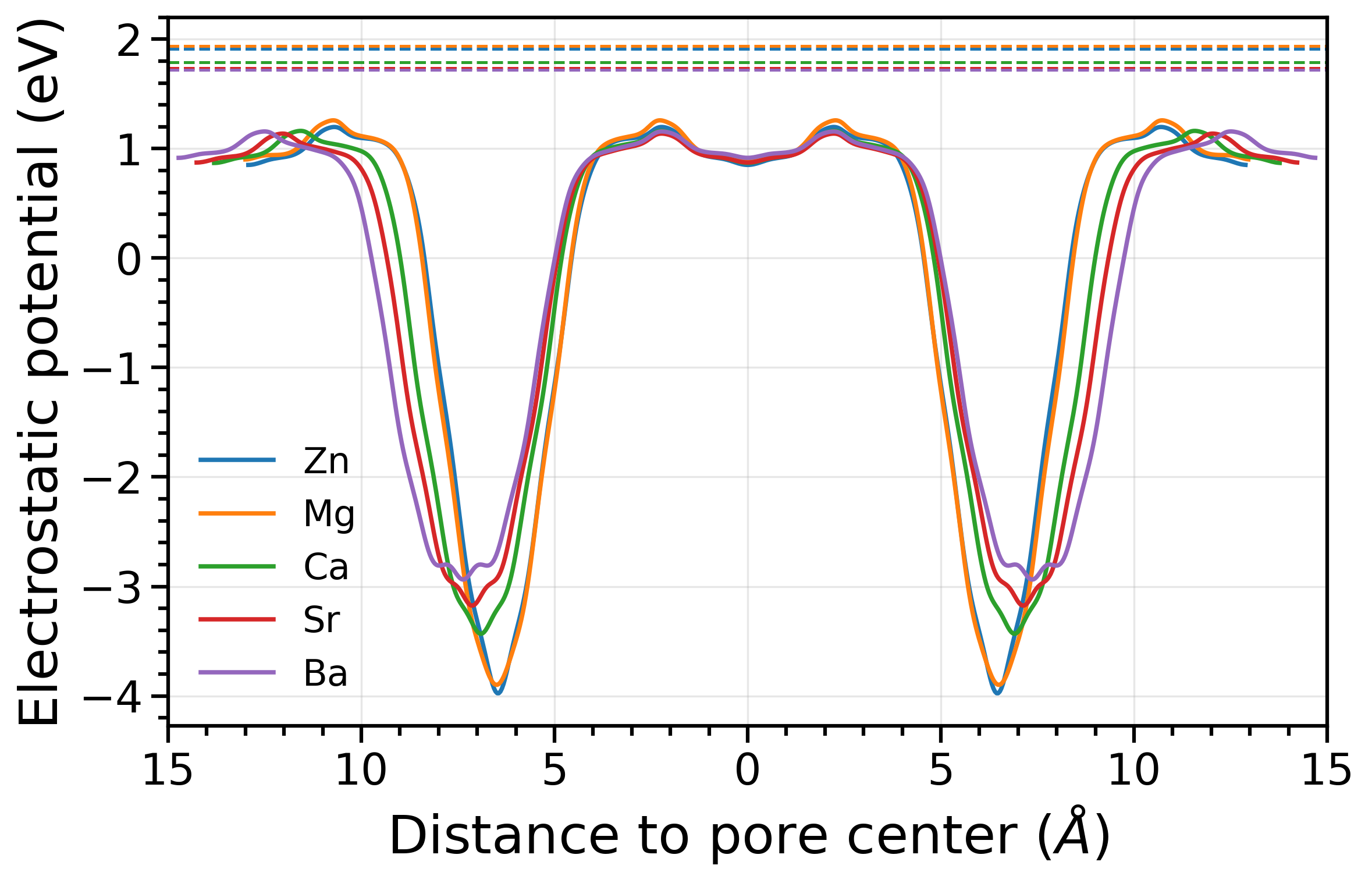}
    \caption{Planar-averaged electrostatic potential profiles (solid lines) as a function of the distance from the pore center along a cartesian direction for pristine MOF-5 (Zn metal node)  and its variants with Mg, Ca, Sr, and Ba. The corresponding spherically averaged electrostatic potentials used as vacuum references are indicated by dashed horizontal lines.} 
    \label{fig:macro_potential_centered}
\end{figure}
We start our analysis by examining the planar-averaged electrostatic potential for unfunctionalized MOF-5 variants (Fig.~\ref{fig:macro_potential_centered}) referenced to the corresponding vacuum levels. Substituting Zn with the other metallic species leaves this reference essentially unchanged, spanning 1.72--1.94~eV across the series (Table~S2), i.e. a variation of at most 0.22~eV. The smooth potential region around the pore center therefore provides a consistent internal reference for comparing the levels of the different frameworks on a common vacuum-energy scale.

\begin{table}[h]
    \centering
    \caption{Calculated fundamental gaps ($E_{\mathrm{g}}$) of pristine MOF-5 as a function of isotropic volumetric strain.}
    \label{tab:strain_bandgaps}
    \begin{tabular}{cc}
        \toprule
        Isotropic Strain (\%) & $E_{\mathrm{g}}$ (eV) \\
        \midrule
        -5.00 & 3.62 \\
        -3.75 & 3.70 \\
        -2.50 & 3.75 \\
        -1.25 & 3.79 \\
         0.00 & 3.83 \\
        +1.25 & 3.86 \\
        +2.50 & 3.89 \\
        +3.75 & 3.92 \\
        +5.00 & 3.91 \\
        \bottomrule
    \end{tabular}
\end{table}

\begin{figure}[t]
    \centering
    \includegraphics[width=\linewidth]{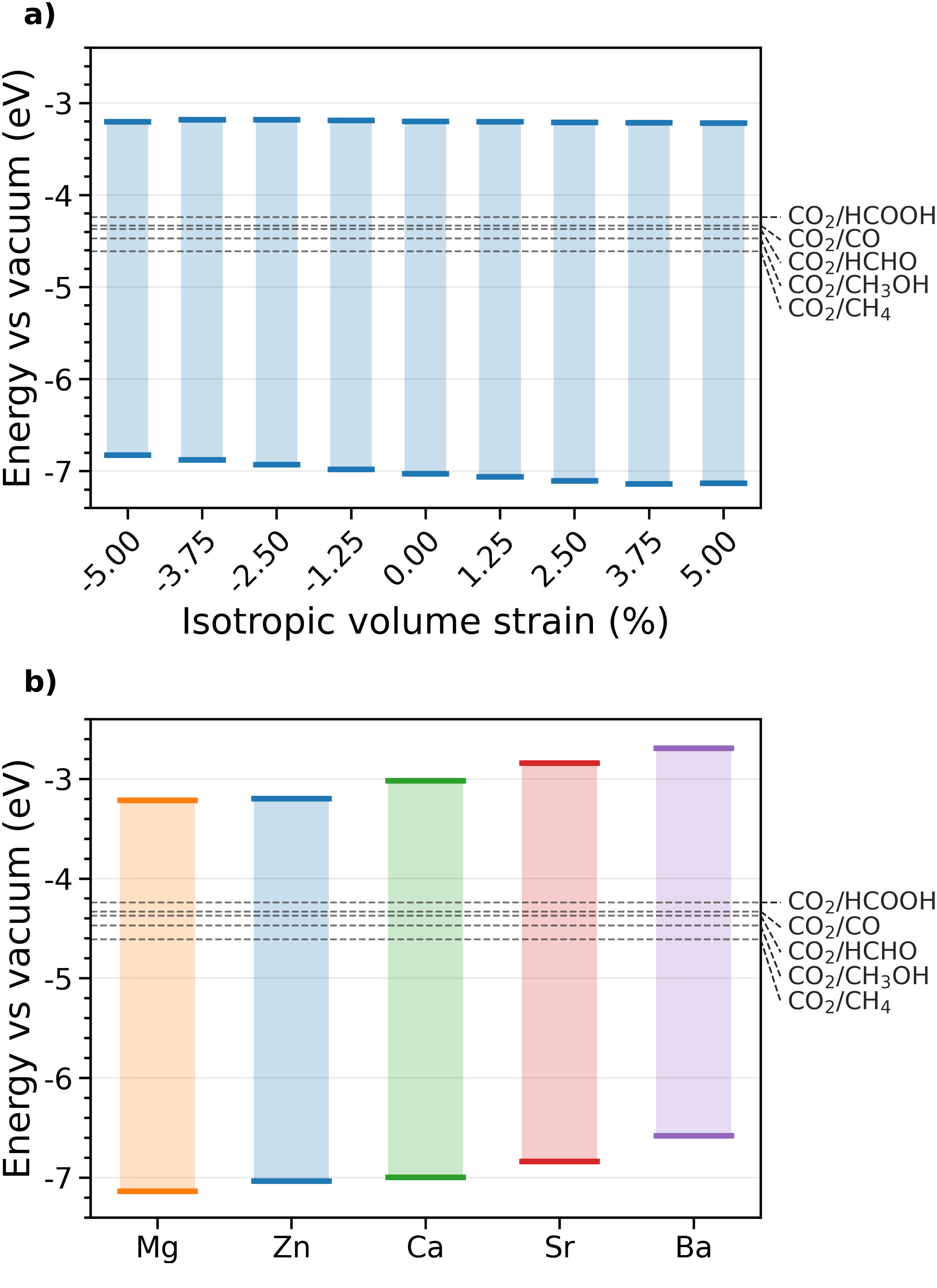}
    \caption{Frontier electronic levels as a function of (a) applied isotropic volume strain from $-5.0$ to $+5.0$\% for pristine MOF-5 and (b) metal-node substitution in MOF-5. All energy levels are referenced to the vacuum level. The dashed horizontal lines denote the vacuum-referenced electron energies corresponding to the aqueous CO$_2$ reduction potentials at pH=0~\cite{kalamaras2018solar}.}
    \label{fig:metal-strain}
\end{figure}

With a common energy scale established, we first evaluate the effects of isotropic volumetric strain ($-5.0\%$ to $+5.0\%$) on the frontier levels of MOF-5. Electrode potentials are referred to the standard hydrogen electrode (SHE = 4.44~$\pm0.02$~V~\cite{trasatti1986absolute,kalamaras2018solar}) and converted to the absolute electron-energy scale using $E_{\mathrm{vac}} = -(4.44~\mathrm{eV} + eE_{\mathrm{SHE}})$. As shown in Fig.~\ref{fig:metal-strain}a, the CBM remains pinned $\sim$1~eV above the CO$_2$ reduction levels across the entire strain window. Consequently, slight variations in the band gap of 0.3~eV (Table~\ref{tab:strain_bandgaps}) are driven almost exclusively by shifts in the VBM, which, however, are insufficient to extend light absorption of MOF-5 into the visible region.

As a second strategy, we substitute the Zn nodes with Mg, Ca, Sr, and Ba atoms. Similar to strain, metal-node exchange induces only negligible variations in the fundamental gaps, which remain within 3.83--4.00~eV across the entire series (Table~\ref{tab:metal_bandgaps}). On the other hand, the presence of different metal cations influences not only the VBM but also the CBM, which remains consistently 1.0--1.6~eV above the CO$_2$ reduction levels, shifting upward with increasing cation size (Fig.~\ref{fig:metal-strain}b). Metal exchange thus preserves favorable reduction alignment with a large excess driving force, but again fails to narrow the electronic gap into the visible range. This weak electronic response is particularly striking given the massive structural expansion across the series. The cubic lattice parameter increases from 25.85~\AA{} in MOF-5 to 29.47~\AA{} in the Ba-containing variant, corresponding to an almost 50\% increase in cell volume (Table~S1). Yet the fundamental gaps of Ba-MOF-5 and Zn-MOF-5 differ by a mere 0.06~eV, \textit{i.e.}, five times less than the 0.30~eV variation produced by $\pm5$\% isotropic strain applied to the pristine framework.

\begin{table}[h]
    \centering
    \caption{Calculated fundamental gaps ($E_{\mathrm{g}}$) of metal-node substituted $\text{M}_4\text{O}$-MOF-5 derivatives ($\text{M} = \text{Zn, Mg, Ca, Sr, Ba}$).}
    \label{tab:metal_bandgaps}
    \begin{tabular}{cc}
        \toprule
        Metal Node & $E_{\mathrm{g}}$ (eV) \\
        \midrule
        Zn & 3.83 \\
        Mg & 3.92 \\
        Ca & 3.98 \\
        Sr & 4.00 \\
        Ba & 3.89 \\
        \bottomrule
    \end{tabular}
\end{table}

\begin{figure*}
\centering
\includegraphics[width=\textwidth]{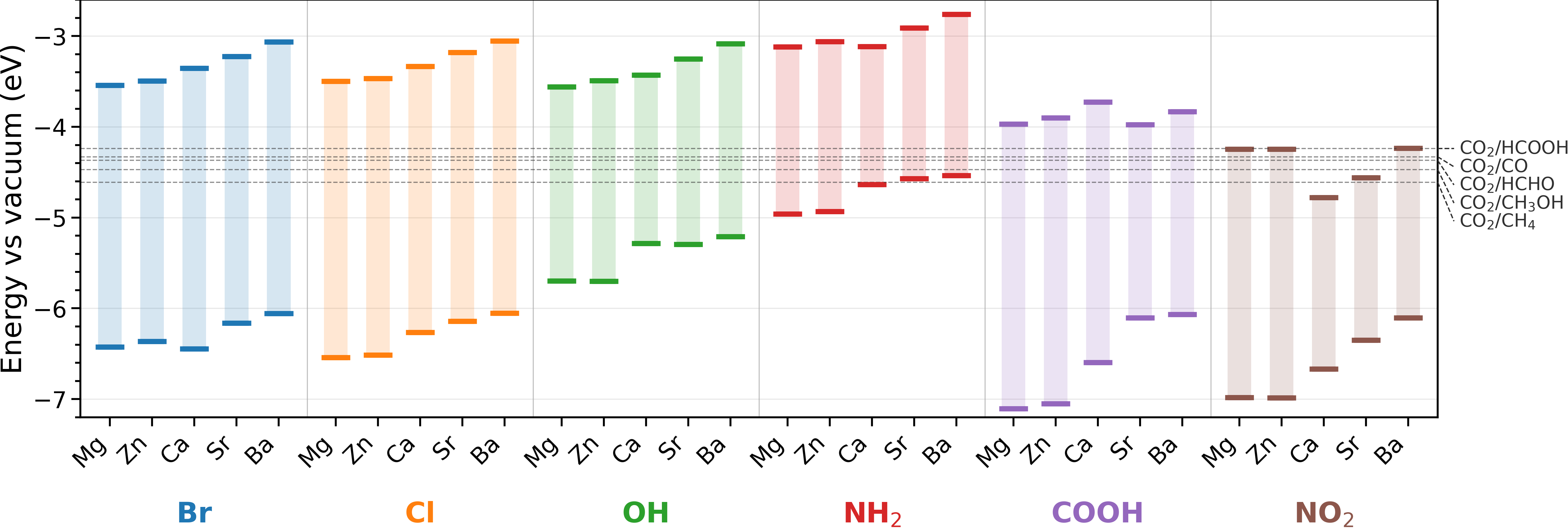}
\caption{Vacuum-aligned VBM and CBM of functionalized MOF-5 variants. Dashed horizontal lines denote the vacuum-referenced electron energies corresponding to the aqueous CO$_2$/HCOOH, CO$_2$/CO, CO$_2$/HCHO, CO$_2$/CH$_3$OH, and CO$_2$/CH$_4$ reduction potentials at pH~=~0~\cite{kalamaras2018solar}.}
\label{fig:functionalization}
\end{figure*}

Given the inability of both strain and metal-node substitution to tune the fundamental gap of MOF-5 to the desired range for CO$_2$ photocatalysis, we turn to linker functionalization as the decisive electronic modulation knob. Decorating the organic linkers with electron-donating and electron-withdrawing groups (Fig.~\ref{fig:mof5-scheme}c) drastically reduces the electronic gaps  (Table~\ref{tab:functionalized_bandgaps}) through the introduction of in-gap states~\cite{Edzards2024} which maintain the CBM well above the target reduction levels (Fig.~\ref{fig:functionalization}). Except for NO$_2$, the CBM is formed by $\pi^*$ states of the BDC linker, while the functional groups shape the top of the valence band: OH and NH$_2$ introduce an orbital localized on the functional group, Br, Cl, and COOH states hybridize with the $\pi$ orbitals of the linker, and in Sr- and Ba-COOH the VBM has O-localized character~\cite{Edzards2024}. The lowest-energy transitions are therefore expected to involve these states and the $\pi^*$ states of the linker.
Introducing the reduction overpotential, $\eta_{\mathrm{red}} = E_{\mathrm{CBM}} - E_{\mathrm{red}}$, with the condition $\eta_{\mathrm{red}}>0$ ensuring a thermodynamically feasible process, we find that across the five target reduction couples, the potentials span a narrow 0.37~eV window, ranging from $-4.24$~eV (CO$_2$/HCOOH) to $-4.61$~eV (CO$_2$/CH$_4$). Consequently, the CO$_2$/HCOOH couple presents the most stringent thermodynamic benchmark: any CBM above this level guarantees sufficient driving force for all five reduction pathways.

Across the halogenated and hydroxylated frameworks, the CBM shifts systematically upward, following the size of the metal ionic radius (Mg~$<$~Zn~$<$~Ca~$<$~Sr~$<$~Ba) with $\eta_{\mathrm{red}}$ spanning $\approx0.4$--$0.5$~eV from Mg to Ba. The NH$_2$-functionalized series follows a similar trend, in contrast to COOH- and NO$_2$-containing frameworks, where the impact of cation size is itself linker-dependent. Among specific functional groups, halogenation (Br, Cl) yields robust reduction overpotentials ($\eta_{\mathrm{red}} = 0.70$--$1.18$~eV for Br; $0.74$--$1.19$~eV for Cl, Fig.~\ref{fig:functionalization}) that increase monotonically from Mg to Ba. However, their fundamental gaps remain comparatively large (2.87--3.09~eV for Br; 2.93--3.05~eV for Cl, Table~\ref{tab:functionalized_bandgaps}).  Relative to the halogens, the OH group raises the VBM by 0.7--1.2~eV while leaving the CBM largely unaffected (Fig.~\ref{fig:functionalization}), due to localized OH-derived states introduced at the valence-band edge~\cite{Edzards2024}. Finally, NH$_2$ functionalization leads to the most pronounced gap narrowing (1.52--1.87~eV, Table~\ref{tab:functionalized_bandgaps}) alongside exceptionally large overpotentials ($\eta_{\mathrm{red}} = 1.12$--$1.48$~eV, Fig.~\ref{fig:functionalization}). Here, both frontier levels shift upward, elevating the VBM to a range between $-4.96$ and $-4.54$~eV.

\begin{table}
    \centering
    \caption{ Calculated fundamental gaps (in eV) of functionalized MOF-5 systems for different metal nodes. }
    \label{tab:functionalized_bandgaps}
    \begin{tabular}{lcccccc}
        \toprule
        Metal &  Br & Cl & OH & NH$_2$ & COOH & NO$_2$ \\
        \midrule
        Mg &  2.89 & 3.04 & 2.14 & 1.84 & 3.14 & 2.73 \\
        Zn &  2.87 & 3.05 & 2.21 & 1.87 & 3.15 & 2.74 \\
        Ca &  3.09 & 2.93 & 1.86 & 1.52 & 2.87 & 1.89 \\
        Sr &  2.94 & 2.96 & 2.04 & 1.66 & 2.13 & 1.79 \\
        Ba &  2.99 & 3.00 & 2.12 & 1.78 & 2.24 & 1.87 \\
        \bottomrule
    \end{tabular}%
    
\end{table}

The COOH functionalization clears all five couples at pH~=~0 but yields the smallest driving forces across the series ($\eta_{\mathrm{red}} = 0.26$--$0.51$~eV, Fig.~\ref{fig:functionalization}). As a strong electron-withdrawing substituent, COOH stabilizes both frontier levels, resulting in modest gap narrowing for Mg, Zn, and Ca (2.87--3.15~eV) and more significant reductions for Sr and Ba (2.13 and 2.24~eV, Table~\ref{tab:functionalized_bandgaps}). This contrast has clear qualitative implications for solar light harvesting: while the wide gaps of the Zn, Mg, and Ca analogs remain largely constrained to the ultraviolet region, the substantial gap reduction induced by Sr- and Ba-nodes markedly improves the potential for visible-light absorption~\cite{Li2018}. Furthermore, the COOH-linker contracts the cubic lattice parameter by 0.6--1.1\% relative to the mean value of the other functionalizations for every metal node (Table~S1), corresponding to a volume contraction of 1.7--3.3\%, consistent with inter-linker hydrogen bonding between neighboring carboxyl groups. 

For the NO$_2$-functionalized series, the CBM lies 0.7--1.4~eV lower than in the corresponding halogenated frameworks (Fig.~\ref{fig:functionalization}), reflecting a low-lying unoccupied state localized on the nitro group~\cite{Edzards2024}. This pronounced stabilization drops Ca-MOF-5-NO$_2$ ($E_{\mathrm{CBM}} = -4.78$~eV) below all target couples (0.54~eV below CO$_2$/HCOOH and 0.17~eV below CO$_2$/CH$_4$), eliminating its thermodynamic viability for electron transfer. Similarly, Sr-MOF-5-NO$_2$ ($E_{\mathrm{CBM}} = -4.56$~eV) clears only the CO$_2$/CH$_4$ couple by 0.05~eV. The CBM of Zn-, Mg- and Ba-MOF-5 with NO$_2$ functionalization all lie within 10~meV of the CO$_2$/HCOOH level (Fig.~\ref{fig:functionalization}). Since these differences are smaller than thermal energy at room temperature and the uncertainty of the adopted alignment procedure, these three systems are treated as effectively degenerate with the CO$_2$/HCOOH threshold.

The primary design objective for solar-driven CO$_2$ catalysis is to balance electron-transfer driving force against kinetic energy losses, rather than simply maximizing $\eta_{\mathrm{red}}$. Excessively large overpotentials dissipate energy through thermalization losses~\cite{Bolton1985,Walter2010}, while an overly elevated CBM accelerates the competing hydrogen-evolution reaction and risks reaching the reductive decomposition limit of the framework~\cite{Gerischer1977}. An efficient catalyst must therefore possess  a CBM safely above the target CO$_2$/product levels without exceeding them by an unnecessarily large margin. At pH~=~0, the halogenated, OH-, NH$_2$-, and COOH-functionalized frameworks meet the first condition, whereas the NO$_2$ series does not. Conversely, NH$_2$- and, to a lesser extent, OH-terminated and halogenated MOF-5 variants overshoot this threshold by providing excessive driving force. For these linkers, the next logical optimization step is to lower $\eta_{\mathrm{red}}$ toward the minimum value required for CO$_2$ reduction, thereby minimizing thermalization losses and maximizing overall photochemical efficiency.


The most elegant and straightforward way to tune $\eta_{\mathrm{red}}$ into a favorable range for a given reduction half-reaction is to adjust the solution pH~\cite{Bolts1976,liu2019ph}. At room temperature, as for any proton-coupled electron-transfer reaction, the redox potentials follow the standard Nernst relation, shifting the reduction levels upward on the absolute scale by 59~meV per pH unit~\cite{kalamaras2018solar}. Treating the MOF band edges as pH-independent yields~\cite{torkashvand2026, pham2014interfacial}:
\begin{equation}
\eta_{\mathrm{red}}(\mathrm{pH}) = \eta_{\mathrm{red}}(0) - 0.059\,\mathrm{pH}.
\label{eq:eta-ph}
\end{equation}

\begin{figure*}
    \centering
    \includegraphics[width=0.875\textwidth]{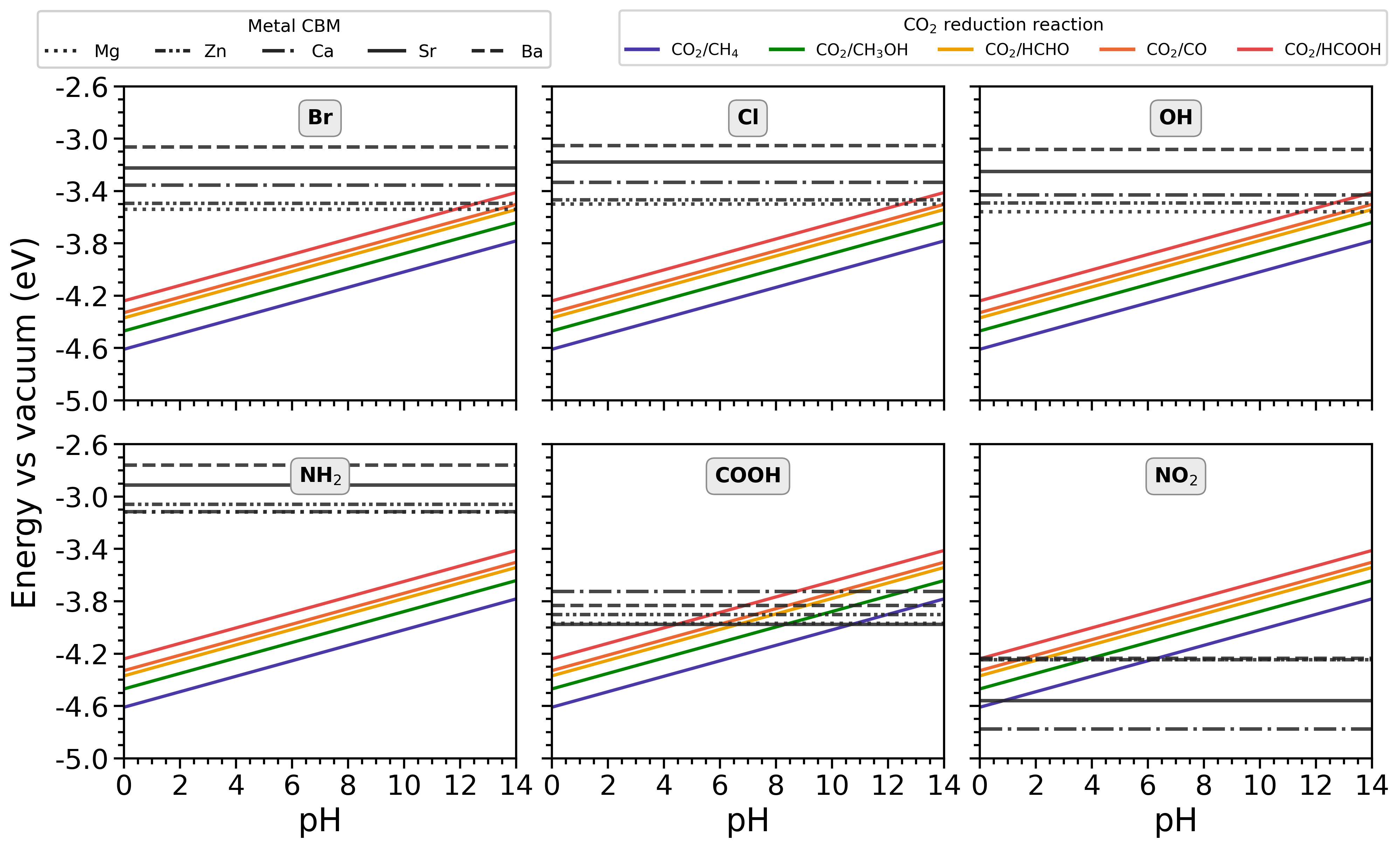}
    \caption{pH dependence of the reduction overpotential for functionalized M--MOF-5 (M~=~Mg, Zn, Ca, Sr, Ba), shown separately for the Br, Cl, OH, NH$_2$, COOH, and NO$_2$ linker functionalizations. The colored diagonal lines are the vacuum-referenced electron energies of the aqueous CO$_2$/HCOOH, CO$_2$/CO, CO$_2$/HCHO, CO$_2$/CH$_3$OH, and CO$_2$/CH$_4$ couples, referenced to pH~0 and converted using the absolute potential of the SHE, $4.44$~V. Horizontal lines denote the CBM of each metal--linker system, distinguished by line style. $\eta_{\mathrm{red}}$ is positive where the CBM lies above the corresponding level.}
    \label{fig:ph-effect}
\end{figure*}

Adopting $0<\eta_{\mathrm{red}}\leq100$~meV as a target driving-force window that limits thermalization losses~\cite{Bolton1985,Walter2010}, Eq.~\eqref{eq:eta-ph} defines a specific pH range for each metal–linker combination, bounded below by $[\eta_{\mathrm{red}}(0)-0.100]/0.059$ and above by the critical limit $\eta_{\mathrm{red}}(0)/0.059$, beyond which thermodynamic alignment is lost. Since every target window is fixed at a width of 1.7~pH units and the five reduction couples maintain constant relative energy offsets (CO$_2$/CO, CO$_2$/HCHO, CO$_2$/CH$_3$OH and CO$_2$/CH$_4$ lying 0.09, 0.13, 0.23 and 0.37~eV below CO$_2$/HCOOH), the five windows for any given framework are identical in width, merely shifted by 1.5, 2.2, 3.9 and 6.3 pH units. The complete map of accessible products is governed entirely by the initial overpotential $\eta_{\mathrm{red}}(0)$ (Fig.~\ref{fig:ph-effect} and Table~S3). Evaluating these relationships within the conventional range ($0\leq\mathrm{pH}\leq14$) reveals that the operational windows for the three least demanding couples (CO$_2$/HCOOH, CO$_2$/CO, and CO$_2$/HCHO)  partially overlap, whereas CO$_2$/CH$_3$OH and CO$_2$/CH$_4$ form cleanly separated $\mathrm{pH}$ regimes. The H$^+$/H$_2$ level ($-4.44$~eV at pH~=~0) lies between the CO$_2$/HCHO and CO$_2$/CH$_3$OH levels and shifts with pH at the same rate, so that frameworks reducing CO$_2$ to HCOOH, CO, or HCHO can also evolve H$_2$; the competition between the two processes is ultimately controlled by kinetics.

For the halogenated and hydroxylated frameworks, the target windows fall toward the alkaline end of the scale, where only the two smallest cations qualify (Fig.~\ref{fig:ph-effect}). Specifically, Mg-MOF-5-Br reaches the target range for CO$_2$/HCOOH at pH~10.2--11.9, for CO$_2$/CO at 11.7--13.4 and for CO$_2$/HCHO from pH~=~12.4. The Zn, Cl, and OH analogs behave similarly, with Mg-MOF-5-OH opening earliest (CO$_2$/HCOOH at pH~=~9.9--11.6) and additionally reaching CO$_2$/CH$_3$OH near the upper pH limit. Ca enters the target range only near this threshold, with CO$_2$/HCOOH turning on at pH~=~13.3, 13.7, and 12.0 with Br-, Cl-, and OH-functionalization, respectively, with the latter also reaching CO$_2$/CO from pH~=~13.6. 

MOF-5 variants with Sr and Ba fail to enter the target window within the accessible range: their large initial overpotentials [$\eta_{\mathrm{red}}(0)$ of 0.99--1.19~eV] yield critical thresholds at pH 17--20 (Table~S3). Since the variation driven by Br-, Cl-, and OH-functionalizations amounts to at most 0.10~eV in $\eta_{\mathrm{red}}(0)$, \textit{i.e.}, within the uncertainty of our alignment procedure, these three linkers are effectively equivalent on the reduction side. Instead, they are distinguished primarily by their fundamental gaps: 2.87--3.09~eV for Br and 2.93--3.05~eV for Cl compared to 1.86--2.21~eV for OH (Table~\ref{tab:functionalized_bandgaps}). Finally, the NH$_2$-terminated series produces the highest overpotentials [$\eta_{\mathrm{red}}(0) = 1.12$--$1.48$~eV], projecting critical thresholds well beyond the accessible pH range. Consequently, while all five reduction half-reactions remain thermodynamically permitted throughout the entire $0\leq\mathrm{pH}\leq14$ window, none can be brought into the chosen $0\text{--}100~\mathrm{meV}$ target range, resulting in unnecessary kinetic thermalization losses across the board.

The COOH-functionalized series stands out as the MOF-5 family for which pH acts as a genuine design lever. Its small initial overpotentials [$\eta_{\mathrm{red}}(0) = 0.26$--$0.51$~eV] place every operational threshold well inside the accessible $0 \le \mathrm{pH} \le 14$ range (Fig.~\ref{fig:ph-effect}). With Sr and Mg cations, the successive product windows span pH values of 2.8--4.6, 4.3--6.1, 5.0--6.8, 6.7--8.5 and 9.0--10.9 for CO$_2$/HCOOH through CO$_2$/CH$_4$, respectively. The Zn- and Ba-containing analogs are shifted upward by 4.0--5.7 to 10.3--12.0 and 5.2--6.9 to 11.5--13.2, respectively. In contrast, Ca-variants require the most alkaline conditions, reaching CO$_2$/CH$_4$ only at the edge of the accessible regime (pH 7.0--8.7 to 13.3--14.0). 

COOH is therefore the only functional group capable of systematically spanning the complete CO$_2$ reduction sequence within an accessible pH range. Across its series, however, the electronic gap varies strongly with the metal node, dropping from 3.15 and 3.14~eV for Zn and Mg to 2.24 and 2.13~eV for Ba and Sr (Table~\ref{tab:functionalized_bandgaps}), even with their reduction overpotential remaining nearly identical. Overall, Sr- and Ba-containing variants combine the favorable, highly tunable reduction energetics of the COOH-functionalized series with a gap well inside the visible range, whereas their Mg- and Zn-based counterparts offer identical pH windows at the cost of ultraviolet-limited gaps.

NO$_2$ sits at the opposite extreme. Its pH~=~0 overpotential for CO$_2$/HCOOH is zero (within 10~meV) for Mg, Zn and Ba, while it assumes negative values for Ca and Sr, restricting the accessible windows to the acidic regime (pH~$<6.3$, Fig.~\ref{fig:ph-effect}). Specifically, Mg-, Zn-, and Ba-containing variants reach the target overpotential range for CO$_2$/CO below pH~1.4--1.6 and for CO$_2$/CH$_4$ at pH~4.4--6.3 (Table~S3). In the presence of Sr, thermodynamic alignment is retained exclusively for CO$_2$/CH$_4$ below pH~0.8, whereas with Ca, the CBM lies 0.54~eV below CO$_2$/HCOOH, failing to align with any reduction couple across the entire accessible interval. Since increasing the pH monotonically reduces $\eta_{\mathrm{red}}$, NO$_2$ offers the narrowest pH range of all considered functionalizations, rendering it effectively unsuitable for $\mathrm{pH}$ tuning of the accessible products.

The pH windows discussed above are best regarded as indicative of the pH regime: an alignment uncertainty of about 0.1~eV corresponds to 1.7 pH units, while a partial co-shift of the band edges with pH, \textit{e.g.}, through linker protonation, would move all windows to higher pH without altering their order. Moreover, MOF-5 degrades upon exposure to water and even moist air~\cite{Kaye2007,greathouse2006role,han2010water}, so that the $0\leq\mathrm{pH}\leq14$ range considered here is to be understood as a formal measure of the proton activity, most relevant to sacrificial-donor configurations in organic or mixed organic--aqueous media~\cite{Pellegrin2017,Takeda2017}.

\section{DISCUSSION} 
The diverse strategies explored to tailor the electronic structure of MOF-5 provide useful guidance for photocatalytic CO$_2$ reduction. Volumetric strain has only a minor effect on the electronic structure and does not bring the absorption edge into the visible range. Similarly, replacing Zn with Mg, Ca, Sr, or Ba is insufficient to achieve visible-light absorption. In contrast, linker functionalization appears as the most effective knob, introducing in-gap states~\cite{Edzards2024} and narrowing the fundamental gap to values compatible with visible-light absorption. Among the considered terminations, Br and Cl bring the gap near the ultraviolet–visible boundary, OH, NH$_2$ and NO$_2$ shift it well into the visible, and COOH spans both regimes depending on the metal node.

A fundamental gap in the visible region alone is, however, insufficient to guarantee appropriate CO$_2$-reduction energetics. At pH~=~0, several functionalized frameworks exhibit unfavorable reduction overpotentials. Tuning the solution pH provides a simple route to modulate $\eta_{\mathrm{red}}$ and thereby control reaction feasibility, with the optimal pH depending on both BDC decoration and the specific metal node. The resulting trends establish a clear hierarchy among the linker functionalizations:
\begin{itemize}
    \item Br, Cl and OH provide only limited accessibility, reaching CO$_2$/HCOOH, CO$_2$/CO and CO$_2$/HCHO under strongly alkaline conditions, for Mg, Zn, and Ca nodes. Since their $\eta_{\mathrm{red}}$ values differ by less than the uncertainty of the alignment procedure, they are effectively equivalent on the reduction side, while differing in their gaps, which favor OH (1.86--2.21~eV). 
\item NH$_2$ is less effective for optimizing driving forces: although all reduction half-reactions remain thermodynamically allowed, they require larger  overpotentials ($\eta_{\mathrm{red}} = 1.12$--$1.48$~eV) that lead to inefficient thermal losses. 
\item COOH is the most favorable functionalization, providing accessible $0<\eta_{\mathrm{red}}\leq100$~meV windows spanning the entire CO$_2$ reduction sequence (from HCOOH to CH$_4$) across the considered pH values. Its performance is particularly strong with Sr and Ba nodes, providing gaps of 2.13 and 2.24~eV, respectively. In contrast, Mg and Zn cations lead to larger gaps in the near-ultraviolet range, and Ca requires highly alkaline conditions. Within the COOH series, the linker fixes the reduction energetics while the metal node determines the fundamental gap, the two acting as largely independent design handles. 
\item NO$_2$ represents the least favorable functionalization overall, with Ca unable to support any reduction pathways, Sr retaining only CO$_2$/CH$_4$ below pH~=~0.8, and Mg, Zn and Ba providing only marginal accessibility in the acidic regime.
\end{itemize}
Ultimately, solution pH serves as an active knob to control which CO$_2$ reduction products are thermodynamically accessible.

In addition to governing fundamental gaps and band alignments, linker functionalization alters the spatial distribution of the frontier states. To evaluate this effect, we consider the spatial overlap of the frontier states as a final electronic descriptor, defined as:
\begin{equation}
S = \int \hat{\rho}_{\mathrm{CBM}}(\mathbf{r}) \hat{\rho}_{\mathrm{VBM}}(\mathbf{r})\,\mathrm{d}\mathbf{r},
\label{eq:overlap}
\end{equation}
where $\hat{\rho}_{\mathrm{CBM}}$ and $\hat{\rho}_{\mathrm{VBM}}$ denote the normalized CBM and VBM densities evaluated at the $\Gamma$ point (Table~\ref{tab:overlap_density}), which are taken as qualitatively reliable proxies for electron and hole distributions. Under this assumption, a smaller $S$ value indicates reduced spatial overlap between the frontier states. 

\begin{table}
\centering
\caption{Spatial overlap $S$ between the normalized VBM and CBM densities ($10^{-3}$~\AA$^{-3}$) in non-functionalized and functionalized MOF-5 structures.}
\label{tab:overlap_density}
\begin{tabularx}{\linewidth}{l*{7}{>{\centering\arraybackslash}X}}
\toprule
Metal & Pristine & Br & Cl & OH & NH$_2$ & COOH & NO$_2$ \\
\midrule
Mg & 1.769 & 2.333 & 0.043 & 0.472 & 0.701 & 0.155 & 0.001 \\
Zn & 1.131 & 0.004 & 0.102 & 0.631 & 0.194 & 0.076 & 0.005 \\
Ca & 3.500 & 1.109 & 0.011 & 0.004 & 0.029 & 0.265 & 0.183 \\
Sr & 1.089 & 0.381 & 0.586 & 0.004 & 0.002 & 0.249 & 0.138 \\
Ba & 0.872 & 0.502 & 0.474 & 0.022 & 0.011 & 0.192 & 0.158 \\
\bottomrule
\end{tabularx}
\end{table}

For non-functionalized MOF-5 structures, $S$ ranges from $0.87\cdot10^{-3}$ to $3.50\cdot10^{-3}$~\AA$^{-3}$. Linker functionalization substantially decreases $S$ for all metal-nodes, with reductions of up to two to three orders of magnitude (Table~\ref{tab:overlap_density}). The sole exception is Mg-MOF-5-Br, where $S$ increases from $1.77\cdot10^{-3}$ to $2.33\cdot10^{-3}$~\AA$^{-3}$, about 32\% above pristine Mg-MOF-5. Overall, organic functionalization strongly enhances the spatial separation of VBM and CBM densities, though both the magnitude and sign of this shift remain sensitive to the specific metal–linker pairing. 

\section{CONCLUSION} 
In summary, we systematically investigated from first principles the effects of isotropic strain, metal-node substitution, linker functionalization, and pH modulation to tune the electronic properties of MOF-5 for photocatalytic CO$_2$ reduction. While volumetric strain and metal-node exchange marginally modify the electronic gap, linker functionalization provides a substantially more effective route, narrowing the fundamental gap down to 1.52~eV through localized in-gap states while simultaneously shifting absolute frontier-level energies. 

The resulting catalytic energetics are governed by a synergistic mechanism: the organic substituent determines the thermodynamic overpotential, while the metal-node choice independently tunes the fundamental gap across a $\sim1$~eV window without sacrificing reduction driving force. While halogenated and hydroxylated frameworks are restricted to initial reduction pathways (HCOOH, CO, HCHO) under alkaline conditions, and NO$_2$ functionalization drives the CBM to thermodynamically unfavorable potentials, the COOH-functionalized series spans the full reduction cascade across the considered pH regimes. Crucially, COOH-decorated frameworks containing Sr and Ba nodes emerge as the premier candidate materials, combining visible-light absorption gaps with thermodynamic access to the complete CO$_2$ reduction sequence. Beyond energy alignment, real-space analysis of the frontier states demonstrates that linker functionalization systematically decreases the frontier-state spatial overlap, which is expected to favor charge separation.

Ultimately, this work demonstrates that decoupling visible-light harvesting from reaction energetics via linker functionalization, metal-node exchange, and pH engineering provides a rational roadmap for designing MOF photocatalysts. This chemically viable and non-disruptive strategy lays the groundwork for rational, computationally guided synthesis and characterization of next-generation solar-fuel materials. Future studies of CO$_2$ adsorption, reaction intermediates, and kinetic barriers on the most promising candidates will be key to establishing the resulting product distribution.

\section*{Author contributions}
\textbf{Julia Santana-Andreo}:  investigation and formal analysis, writing original draft; \textbf{Joshua Edzards:} investigation and formal analysis, review \& editing; \textbf{Surender Kumar:} conceptualization, investigation and formal analysis,  writing original draft;  \textbf{Caterina Cocchi:} conceptualization, resources, supervision, writing – review \&  editing. 
\section*{Conflicts of interest}
There are no conflicts to declare.

\section*{Data availability}
Data of all the ab initio calculations performed in this work are available free of charge in Zenodo at \href{https://doi.org/10.5281/zenodo.22211415}{10.5281/zenodo.22211415} (Record: 22211415). 

\section*{Acknowledgements}
J.S.A. acknowledges funding from the Evonik Stiftung.
J.E. appreciates financial support from the Nagelschneider Stiftung.
This work was partly funded by the German Federal Ministry of Education and Research (Professorinnenprogramm III) and from the State of Lower Saxony (Professorinnen f\"ur Niedersachsen).
The computational resources were provided by the North-German Supercomputing Alliance (NHR), project nic00084.



\balance


\bibliography{main} 
\bibliographystyle{rsc} 
\end{document}